\documentclass[3p,times,twocolumn,sort&compress]{elsarticle}

\usepackage{ecrc}
\volume{00}
\firstpage{1}

\journalname{Nuclear Physics B Proceedings Supplement}
\runauth{L.~Bhattacharya, R.~Ryblewski, and M.~Strickland }
\jid{nppp}
\jnltitlelogo{Nuclear Physics B Proceedings Supplement}
\usepackage{amssymb}
\usepackage[figuresright]{rotating}
\usepackage{lineno}

\usepackage[usenames,dvipsnames]{color}
\definecolor{darkblue}{RGB}{0,0,196}
\usepackage[colorlinks=true,linkcolor=darkblue,citecolor=darkblue,urlcolor=darkblue]{hyperref}

\begin{document}


\begin{frontmatter}

\dochead{}

\title{Photon and dilepton production from a non-equilibrium quark-gluon plasma}

\author[a]{Lusaka Bhattacharya}
\author[b]{Radoslaw Ryblewski}
\author[a]{Michael Strickland}

\address[a]{Department of Physics, Kent State University, Kent, OH 44242 United States}
\address[b]{The H. Niewodnicza\'nski Institute of Nuclear Physics, Polish Academy of Sciences, PL-31342 Krak\'ow, Poland}

\begin{abstract}
We calculate leading-order medium photon and dilepton yields from a quark-gluon plasma using (3+1)-dimensional anisotropic hydrodynamics. In the case of dileptons, at leading-order it is sufficient to take into account non-equilibrium corrections to the rate through the use of anisotropic distribution functions.  In the case of photons, non-equilibrium corrections to the rate are taken into account using a self-consistent modification of the distribution functions and the corresponding anisotropic hard-loop quark self-energies.  We present predictions for the high-energy photon and dilepton spectrum and the photon elliptic flow as a function of invariant mass, transverse momentum, shear viscosity, and initial momentum-space anisotropy.   Our findings indicate that both high-energy photon and dilepton production are sensitive to the assumed level of initial momentum-space anisotropy of the quark-gluon plasma.  In addition, we find that the photon elliptic flow depends on the initial momentum-space anisotropy.
\end{abstract}

\begin{keyword}
Quark-gluon plasma, Electromagnetic emissions, Non-equilibrium dynamics
\end{keyword}

\end{frontmatter}


\section{Introduction}
\label{sec:intro}

Relativistic heavy-ion collisions at Brookhaven National Laboratory's Relativistic Heavy Ion Collider (RHIC) and at CERN's Large Hadron Collider (LHC) are being used to produce and study the properties of a deconfined plasma of quarks and gluons called the quark-gluon plasma (QGP).  One outstanding question concerning the generation of a QGP in heavy-ion collisions concerns the time scale for the thermalization and isotropization.  Theoretical calculations using both weak- and strong-coupling methods find that the QGP created immediately after the initial nuclear impact ($\tau \sim 0.2$ fm/c) is highly anisotropic in local rest frame (LRF) momentum \cite{Strickland:2014pga}.  In order to confirm the existence of such momentum-space anisotropies, it is desirable to have experimental observables that are sensitive the early-time dynamics of the QGP.

In this respect, electromagnetic emissions are an ideal observable since dileptons and photons are weakly coupled to the plasma ($\alpha \ll \alpha_s$) and escape to the detectors with minimal interaction.  In this proceedings contribution, we discuss two recent papers in which we computed dilepton \cite{Ryblewski:2015hea} and photon \cite{Bhattacharya:2015ada} production from a deconfined QGP, taking into account the (3+1)D non-equilibrium evolution of the QGP using anisotropic hydrodynamics \cite{Martinez:2010sc,Florkowski:2010cf,Bazow:2013ifa,Tinti:2014yya,Strickland:2014pga}.  For dilepton and photon production, the work presented herein is an extension of previous studies performed in Refs.~\cite{Mauricio:2007vz,Martinez:2008di,Martinez:2008mc} and Refs.~\cite{Schenke:2006yp,Schenke:2008hw}, respectively, to include a realistic (3+1)D model of the background evolution.

\section{Methodology}

\subsection{Dilepton Rate}

At leading order in the electromagnetic coupling, the dilepton emission rate follows from
\begin{eqnarray}
\frac{d R^{l^+l^-}}{d^4\!P} &=& \int_{\bf p_1}\int_{\bf p_2} \, f_q({\bf p}_1)\,f_{\bar{q}}({\bf p}_2)\, \nonumber \\ 
&& \hspace{5mm} \times\, 
v_{q\bar{q}}\,\sigma^{l^+l^-}_{q\bar{q}}\, \delta^{(4)}(P^\mu-p_1^\mu-p_2^\mu) \, ,
\label{kineticrate}
\end{eqnarray}
where $\int_{\bf p} \equiv \int d^3{\bf p}/(2\pi)^3$, $f_{q ({\bar q})}$ is the phase-space distribution function of quarks (anti-quarks), $\it{v}_{q\bar{q}}$ is the relative velocity between the quark and the anti-quark and $\sigma^{l^+l^-}_{q\bar{q}}$ is the total cross section for the leading-order quark--anti-quark annihilation process, $q + \bar{q} \to \gamma^\ast \to l^+ + l^-$
\begin{equation}
\sigma^{l^+l^-}_{q\bar{q}} = \frac{4\pi}{3} \frac{\alpha^2}{M^2} 
		\left(1 + \frac{2 m_l^2}{M^2}\right) 
		\left(1 - \frac{4 m_l^2}{M^2}\right)^{1/2} .
\label{crosssection}
\end{equation}
For high-energy dilepton pairs, the invariant energies are much greater than the lepton masses, $M \gg m_l$, and one can take $m_l = 0$ above.

In order to take into account the largest corrections to ideal hydrodynamic behavior which arise from the rapid longitudinal expansion of the plasma at early times, the one-particle distribution function for the quarks and anti-quarks can be described at leading order by the following spheroidal form \cite{Romatschke:2003ms}
\begin{equation}
f_{q ({\bar q})}({\bf p},\xi,\Lambda)\equiv f^{\rm iso}_{q ({\bar 
q})}\left(\!\sqrt{{\bf p}^2+\xi({\bf p\cdot \hat{n}})^2},\Lambda \right) ,
\label{distansatz}
\end{equation}
where $\Lambda$ is a transverse-momentum scale and \mbox{$-1 < \xi < \infty$} is the momentum-space anisotropy parameter. We take $f^{\rm iso}_{q ({\bar q})}$ to be a Fermi-Dirac distribution function and $f^{\rm iso}_{g}$ to be a Bose-Einstein distribution.  For details concerning the evaluation of the integral (\ref{kineticrate}), we refer the reader to Refs.~\cite{Ryblewski:2015hea,Martinez:2008di}. 

\subsection{Photon Rate}

To calculate the leading-order photon rate, one must separately compute the rates stemming from hard and soft momentum exchanges due to the presence of an infrared divergence in the rate.  The total rate, which is a sum over the hard and soft contributions, is finite.  For the hard contributions, one can simply compute the Feynman diagrams corresponding to the annihilation and Compton processes.  The rate of photon production due to in-medium quark annihilation and Compton processes can be expressed as 
\begin{eqnarray}
&& \hspace{-1cm} E\frac{dR_{\rm ann}}{d^3q} = 64\pi^3{e_q}^2\alpha_s \alpha \int_{\bf k_1} 
\frac{f_q({\bf k}_1)}{k_1} \int_{\bf k_2}\frac{f_q({\bf k}_2)}{k_2} \nonumber 
\\ && \hspace{-1cm} \times  \int_{\bf k_3} \frac{1+f_g({\bf k}_3)}{k_3} 
\delta^4 (K_1+ K_2- Q- K_3) \left[\frac{u}{t}+\frac{t}{u}\right] ,
\label{photon_rate_ann1}
\end{eqnarray}
and
\begin{eqnarray}
&& \hspace{-1cm} E\frac{dR_{\rm com}}{d^3q}=-128\pi^3 e_q^2\alpha_s \alpha \,
\int_{\bf k_1} \frac{f_q({\bf k}_1)}{k_1} \int_{\bf k_2}\frac{f_g({\bf k}_2)}{k_2} 
 \nonumber\\
&& \hspace{-1cm} \times \int_{\bf k_3} \frac{1-f_q({\bf k}_3)}{k_3} \delta^4 (K_1+ K_2- Q- K_3) \left[\frac{s}{t}+\frac{t}{s}\right], 
\end{eqnarray}
respectively, where $s \equiv (K_1 + K_2)^2$, $t = (K_1 - Q)^2$, $u = (K_2 - Q)^2$, ${e_q}^2 = 2/3$, and 
$f_{q,g}$ are the in-medium quark and gluon distribution functions.  Due to the infrared divergences in these integrals, we first change variables in the first integration to $P \equiv K_1 - Q$ and introduce an IR cutoff $p^*$ on the integration over the exchanged three-momentum {\bf p}. 

Using the Keldysh formalism, the soft contribution can be computed from the trace of the $(12)$ component of the photon polarization tensor
\begin{equation}
E\frac{dR_{\rm soft}}{d^3q}=\frac{i}{2(2\pi)^3}\left(\Pi_{12}\right)^\mu_{~\mu}\!(Q) \, .
\label{soft_photon_rate}
\end{equation}
We evaluate $\left(\Pi_{12}\right)^\mu_{~\mu}$ using the hard-loop resummed
fermion propagator \cite{Schenke:2006fz} which gives a retarded
fermion self-energy of the form
\begin{equation}
\Sigma (P) =\frac{C_F}{4}g^2 \int_{\bf k} \frac{f(\bf k)}{|{\bf k}|}
\frac{K \cdot \gamma}{K \cdot P} \, ,
\label{sigmap}
\end{equation}
where $f({\bf k})\equiv 2(f_q({\bf k})+f_{\bar q}({\bf k}))+4f_g({\bf k})$.  Taking the hard-loop limit where appropriate, one finds 
\begin{equation}
i\left(\Pi_{12}\right)^\mu_{~\mu}\!(Q)= - e^2{e_q}^2N_c 
\frac{8f_{c}({\bf q})}{q}\int_{\bf p}Q_{\nu}{\tilde \Lambda}^{\nu} ({\bf p}) \, ,
\label{pimunu}
\end{equation}
where ${\tilde\Lambda}^{\nu} ({\bf p})= [ {\Lambda^{\nu\alpha}}_\alpha - {\Lambda_\alpha}^{\nu\alpha}+ {\Lambda_\alpha}^{\alpha\nu}]_{p_0=p(\hat {\bf p} \cdot \hat {\bf q})}$, and 
\begin{equation}
\hspace{-3mm}
\Lambda_{\alpha \beta\gamma}(P)=\frac{P_{\alpha}-\Sigma_{\alpha}(P)}
{(P-\Sigma(P))^2} \, {\rm Im}[\Sigma_{\beta}(P)] \, \frac{P_{\gamma}-\Sigma_\gamma^*(P)}
{(P-{\Sigma^*(P)})^2} \, .
\label{soft6}
\end{equation}
The ultraviolet divergence in the integral (\ref{pimunu}) is regulated by a UV cutoff $p^*$ on the length of the three-momentum.

\subsection{QGP evolution}

We assume that the QGP evolves through a non-equilibrium state and that the quark and anti-quark one-particle distribution functions are well approximated by Eq.~(\ref{distansatz}) at all times~\cite{Strickland:2014pga}.  For all results presented herein, we held the final particle multiplicity fixed as we varied $\eta/s$. For details concerning the anisotropic hydrodynamics setup and initial conditions used for the results presented herein, we refer the reader to the original papers \cite{Ryblewski:2015hea,Bhattacharya:2015ada}.

\begin{figure*}[t]
\centerline{
\includegraphics[width=0.48\linewidth]{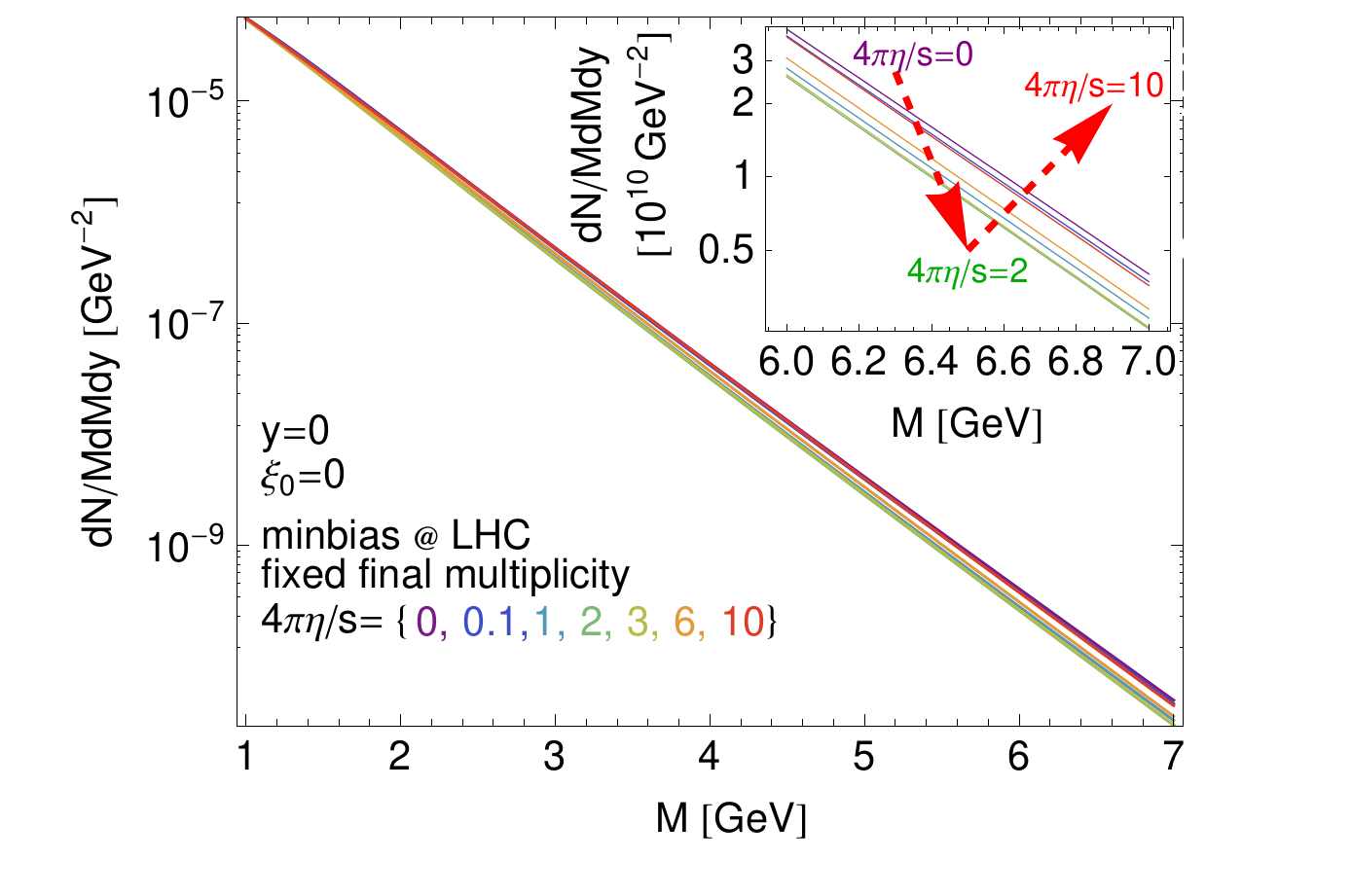}
\includegraphics[width=0.48\linewidth]{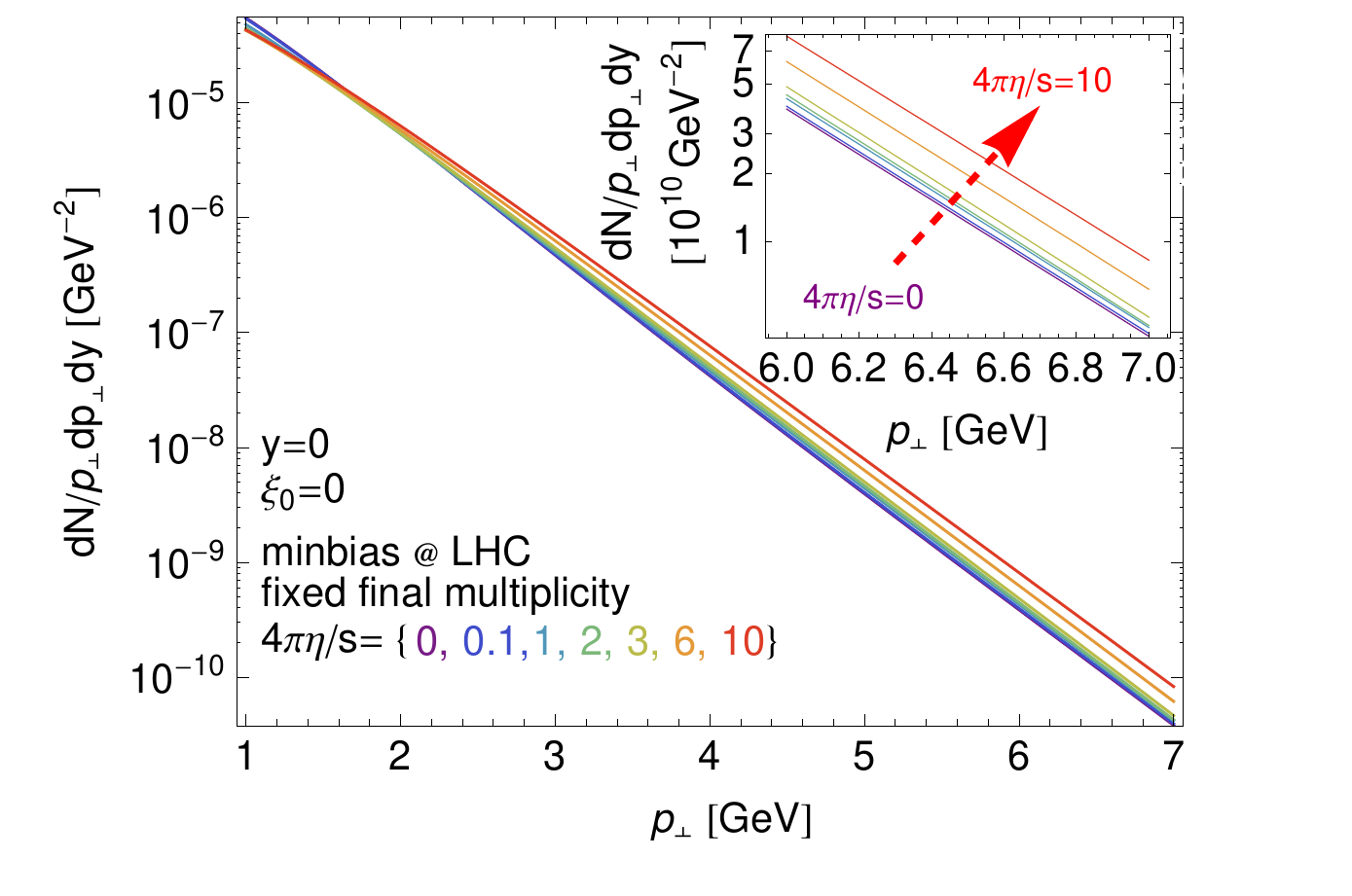}
}
\vspace{-5mm}
\caption{The invariant mass spectra (left) and transverse momentum spectra (right) of dilepton pairs at midrapidity, $y = 0$, for $4 \pi \eta/s \in \{0, 0.1, 1, 2, 3, 6, 10\}$.}
\label{spectrafixfinal}
\end{figure*}

\begin{figure*}[t]
\centerline{
\includegraphics[width=0.48\linewidth]{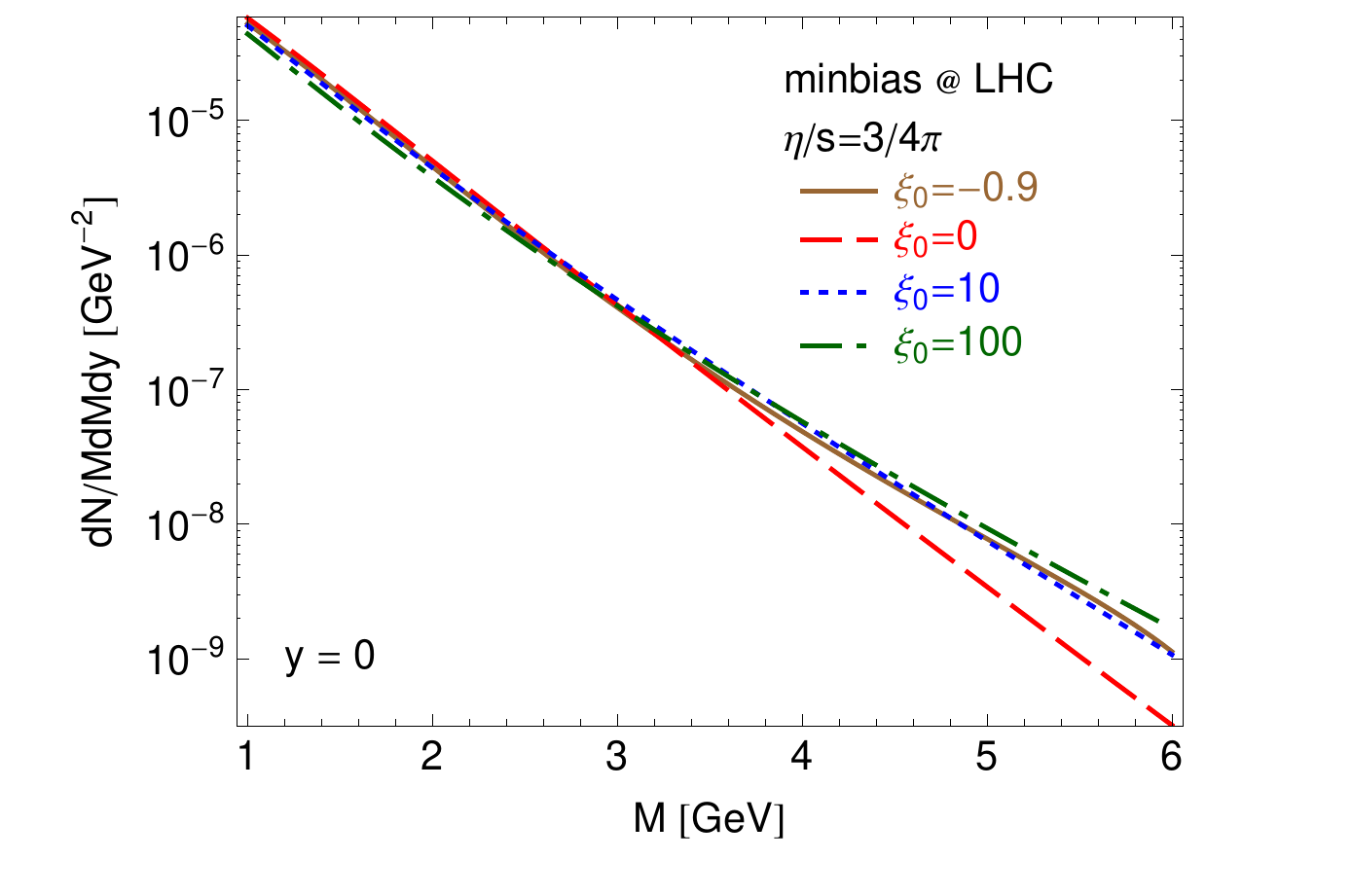}
\includegraphics[width=0.48\linewidth]{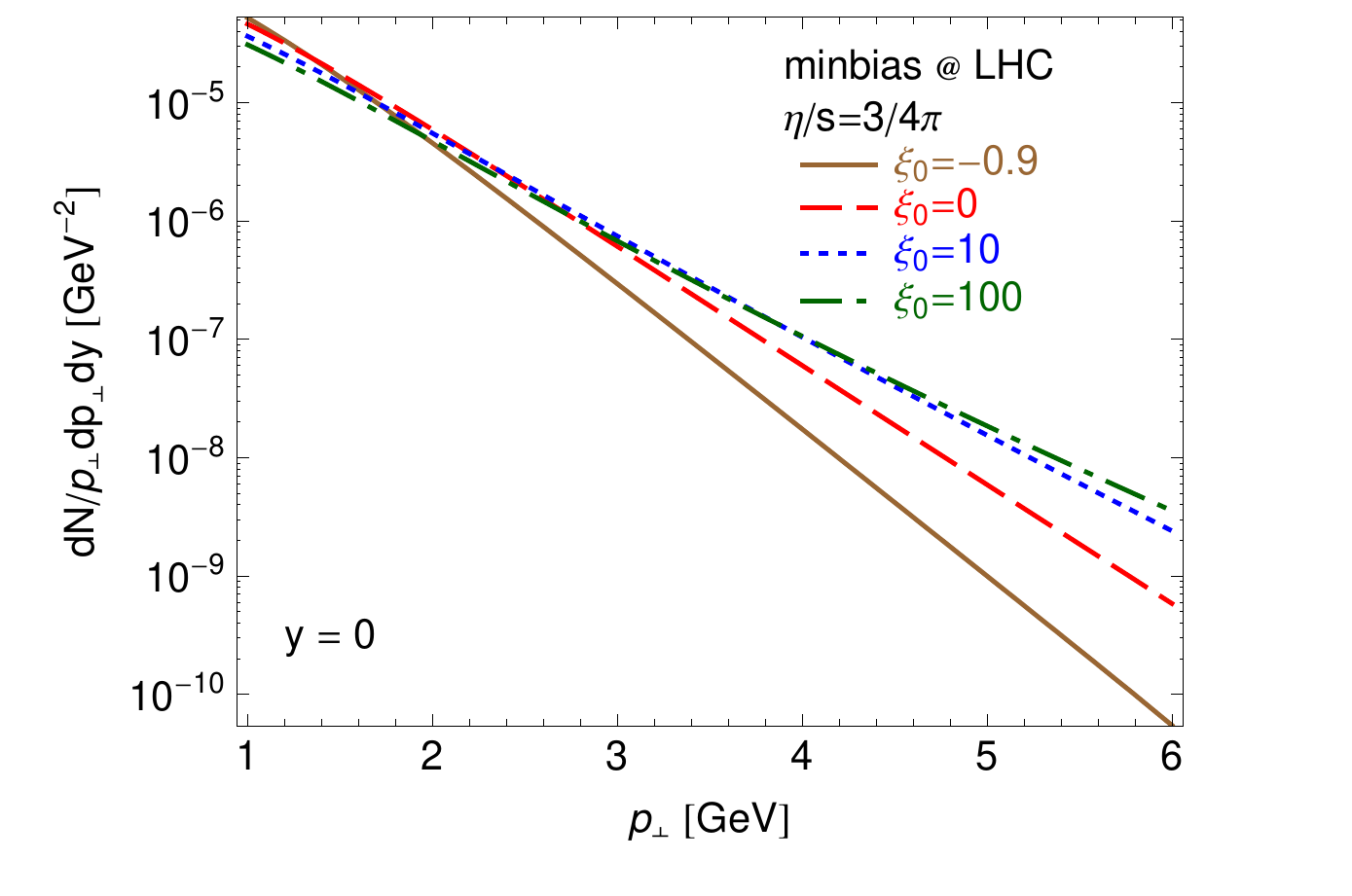}
}
\vspace{-5mm}
\caption{The invariant mass spectra (left) and transverse momentum spectra (right) of dilepton pairs at midrapidity, $y = 0$, for various initial anisotropy conditions, and $4 \pi \eta/s = 3$. The results with $\xi_0 = -0.9, 0, 10$, and $100$ are denoted by brown solid, red dashed, blue dotted and green dot-dashed lines, respectively.}
\label{spectravarxi2}
\end{figure*}

\begin{figure*}[t]
\hspace{2mm}
\includegraphics[width=0.34\linewidth]{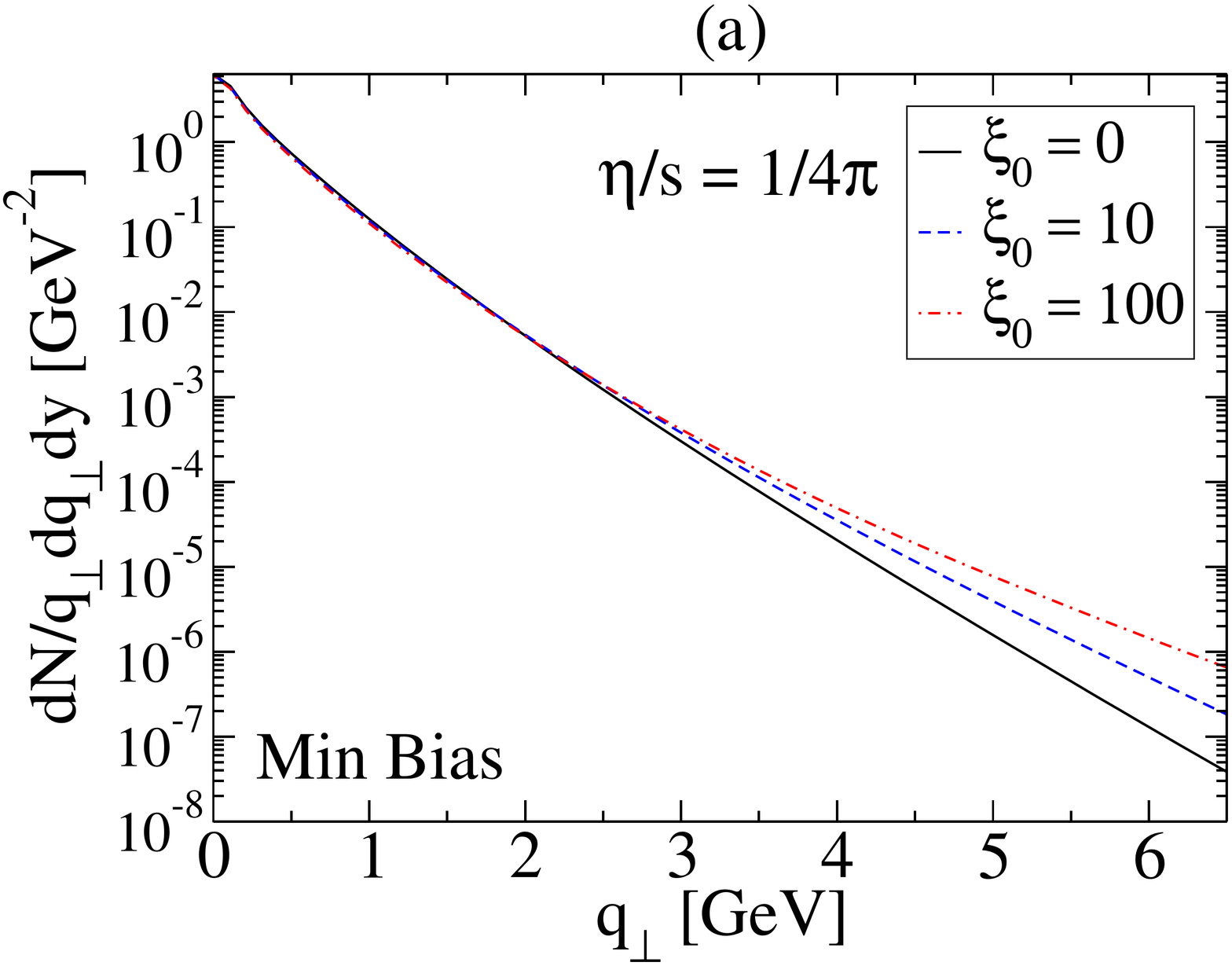}\hspace{-3mm}
\includegraphics[width=0.34\linewidth]{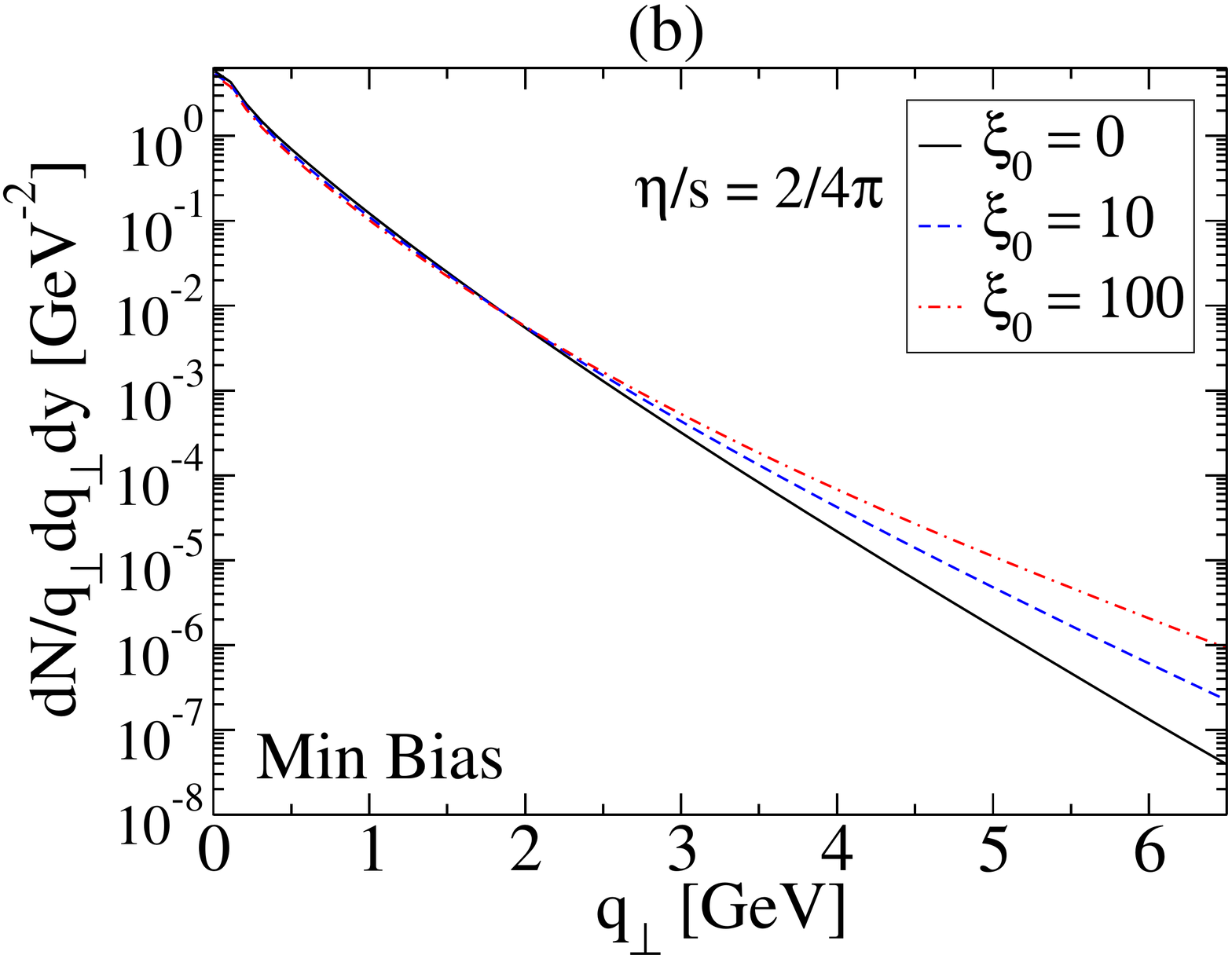}\hspace{-3mm}
\includegraphics[width=0.34\linewidth]{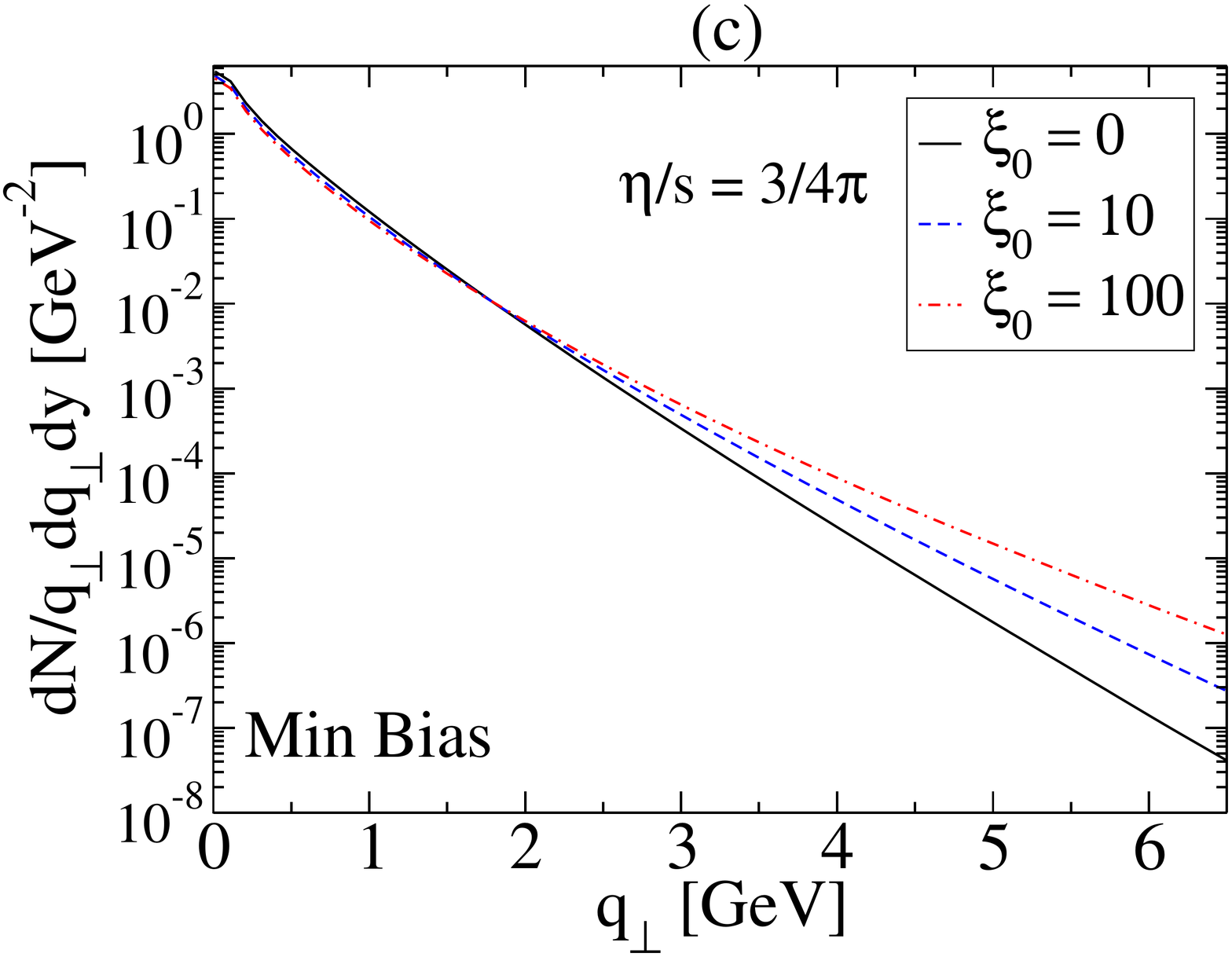}
\vspace{-6.5mm}
\caption{Medium photon spectrum for three different values of viscosity (a) $4\pi\eta/s = 1$ , (b) $4\pi\eta/s = 2$ and (c) $4\pi\eta/s = 3$.  In each panel, the lines correspond to three different values for the 
initial anisotropy in the system $\xi_0 = 0, 10,$ and 100.}
\label{spectrafixmul2}
\end{figure*}

\begin{figure*}[t]
\hspace{2mm}
\includegraphics[width=0.34\linewidth]{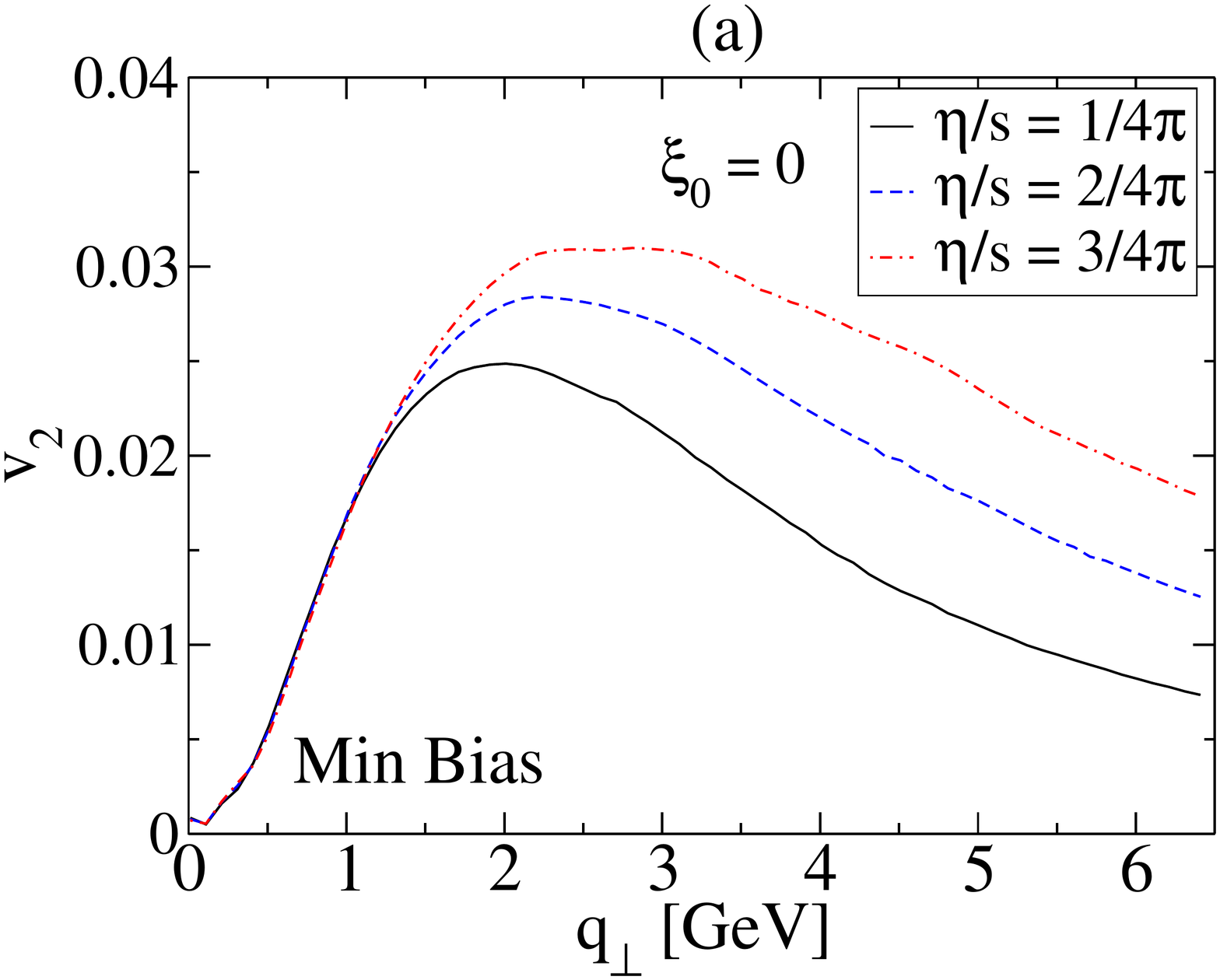}\hspace{-3mm}
\includegraphics[width=0.34\linewidth]{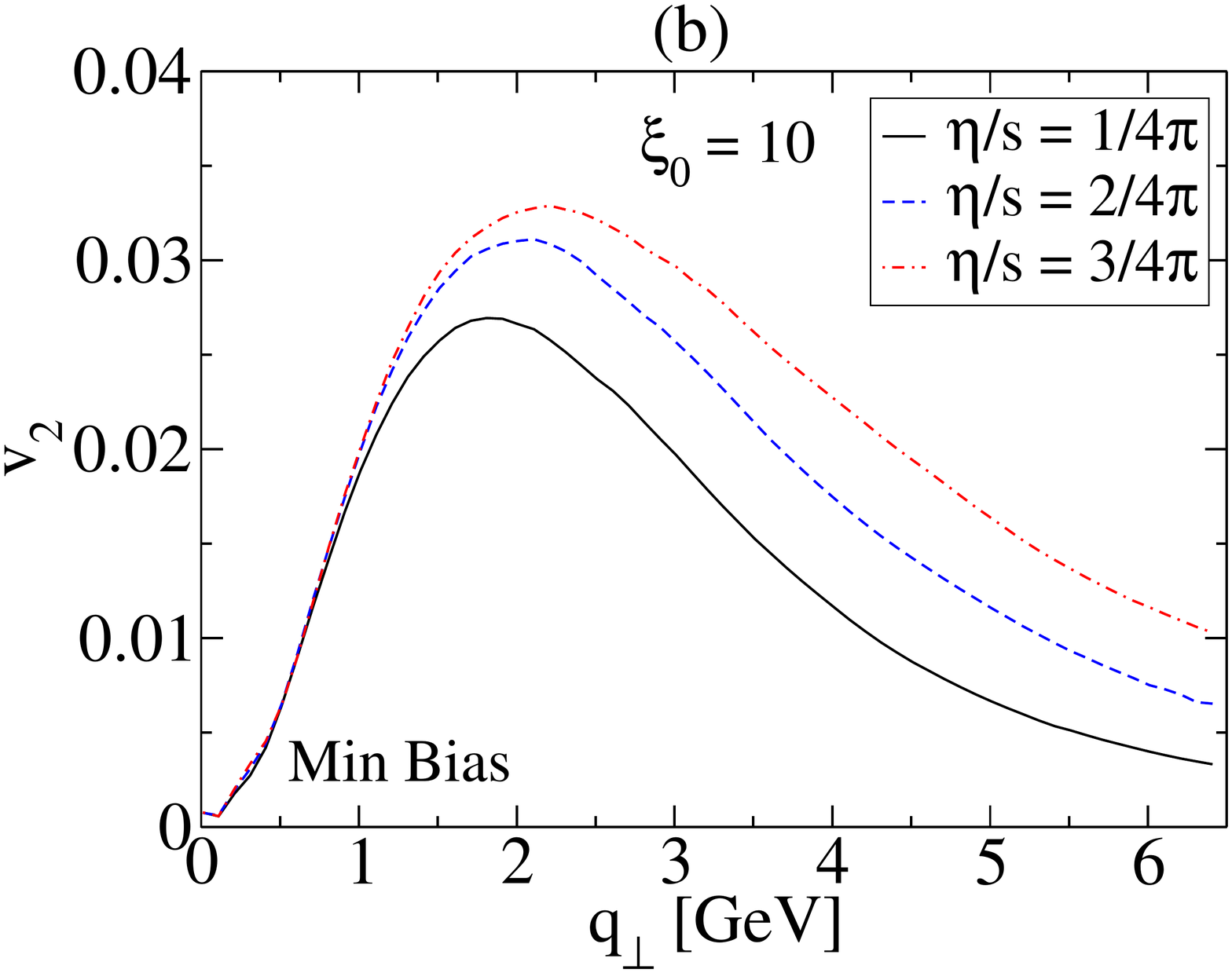}\hspace{-3mm}
\includegraphics[width=0.34\linewidth]{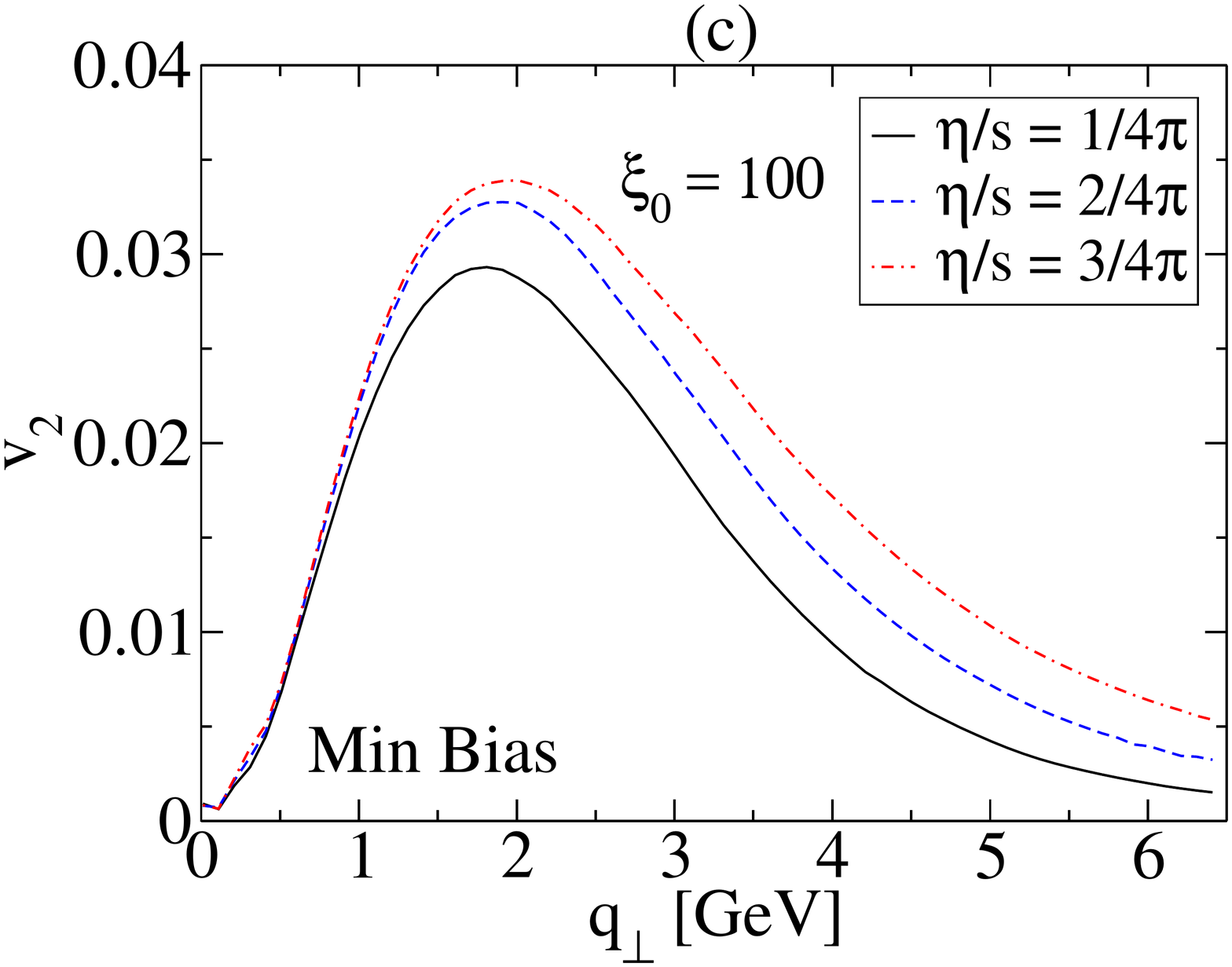}
\vspace{-6.5mm}
\caption{Elliptic flow coefficient $v_2(q_\perp,y=0)$ for three different values of initial anisotropy 
parameter: (a) $\xi_0 = 0$, (b) $\xi_0 = 10$, and (c) $\xi_0 = 100$.  In each panel, the lines correspond to three different values for the shear viscosity to entropy density ratio $4\pi \eta/s =$ 1, 2, and 3.}
\label{V2_2}
\end{figure*}

\section{Results and Conclusions}
\label{sec:results}

In Fig.~\ref{spectrafixfinal}, we present the invariant mass (left) and transverse momentum spectra (right) of dilepton pairs.   From these panels, we see that the spectra do not depend significantly on $\eta/s$.  We observe that the spectra do not necessarily have a monotonic dependence on $\eta/s$ (see e.g. the inset in the left panel of Fig.~\ref{spectrafixfinal}).  This is due to the fact that, for fixed initial temperature, the particle multiplicity is not a monotonically increasing function of $\eta/s$ within anisotropic hydrodynamics (see Fig.~9 of Ref.~\cite{Ryblewski:2015hea} and the surrounding discussion). 

In Fig.~\ref{spectravarxi2}, we present the invariant mass (left) and transverse momentum spectra (right) for dilepton pairs for four different anisotropic initial conditions.  From this figure, we see that the transverse momentum spectra are quite sensitive to the assumed initial momentum anisotropy. For an initially oblate configuration, the spectra are become flatter. The opposite behavior is observed for an initially prolate configuration. This effect is particularly significant for large values of $p_\perp$. The behavior of the invariant mass spectra, on the other hand, is more difficult to understand since, in this case, both oblate and prolate initial conditions lead to a flattening of the spectra.

In Fig.~\ref{spectrafixmul2} we present our results for the photon spectra.  As can be seen from Fig.~\ref{spectrafixmul2}, there is significant variation in the high-energy photon spectrum as one changes the initial anisotropy of the QGP.  For $4\pi\eta/s = 3$, at \mbox{$q_\perp = 6$ GeV}, one finds that the QGP photon spectrum varies by approximately an order of magnitude when varying the initial anisotropy in the range shown.  We note additionally that for $q_\perp \lesssim$ 2 GeV, we see very little effect from varying the initial anisotropy.  In the original paper \cite{Bhattacharya:2015ada}, we found that this conclusion holds for all values of $4\pi\eta/s \gtrsim 1$ with the variation increasing with increasing $\eta/s$.

Finally, in Fig.~\ref{V2_2}, we show our results for the photon elliptic flow, $v_2$.  As can be seen from Fig.~\ref{V2_2}, regardless of the assumed initial momentum-space anisotropy, increasing the shear viscosity of the QGP results in an increase in photon $v_2$.  For the values of $\eta/s$ shown, we see at most a 300\% increase in the photon $v_2$.  Comparing the three panels of Fig.~\ref{V2_2} we also see that, for fixed $\eta/s$, increasing the initial momentum-space anisotropy also results in an increase in the peak photon $v_2$, however, as $\xi_0$ increases there is also a reduction in the photon $v_2$ at large transverse momentum.

In closing, the results of Refs.~\cite{Ryblewski:2015hea} and \cite{Bhattacharya:2015ada}, which are summarized in this proceedings contribution, indicate that both high-energy photon and dilepton production are sensitive to the assumed level of initial momentum-space anisotropy of the QGP.  In addition, we find that the photon elliptic flow depends on the assumed level of initial momentum-space anisotropy of the QGP.  Looking forward, it will be important to include non-spheroidal corrections in the background evolution in order to assess their impact on photon elliptic flow and the photon and lepton spectra.

\vspace{3mm}
\noindent
{\em Acknowledgements}: L.~Bhattacharya was supported by a M.~Hildred Blewett Fellowship from the American Physical Society.  R. Ryblewski was supported by the Polish National Science Center Grant No.~DEC-2012/07/D/ST2/02125.  M.~Strickland was supported by the U.S.~Department of Energy under Award No.~DE-SC0013470.  

\vspace{-3mm}

\bibliographystyle{elsarticle-num}
\bibliography{Strickland_M.bib}

\end{document}